\documentstyle[12pt]{article}\textheight 230mm\textwidth 150mm
            \newcommand{\be}{\begin{eqnarray}}
            \newcommand{\ee}{\end{eqnarray}}
           \newcommand{\eel}[1]{\label{#1}\end{eqnarray}}
\newcommand{\e}[1]{\label{e:#1}\end{eqnarray}}
     
            \newcommand{\ie}{{\em i.e.\ }}
            \newcommand{\ga}{{\gamma}}
 
            \newcommand{\la}{{\lambda}}
            
            \newcommand{\del}{{\delta}}
 \newcommand{\om}{{\omega}}

\newcommand{\bx}{\bar{x}}

\newcommand{\dx}{\dot{x}}

           \newcommand{\ra}{{\rightarrow}}

 \newcommand{\tca}{\tilde{{\cal C}}}
\newcommand{\tphi}{\tilde{{\Phi}}}

\newcommand{\tx}{\tilde{x}}

\newcommand{\cS}{{\cal S}}
\newcommand{\cM}{{\cal M}}

\newcommand{\cC}{{\cal C}}

            \newcommand{\beq}{\begin{quote}}
            \newcommand{\eq}{\end{quote}}
            
            \newcommand{\al}{\alpha}
            \newcommand{\ben}{\begin{enumerate}}
            \newcommand{\een}{\end{enumerate}}
            \newcommand{\bit}{\begin{itemize}}
            \newcommand{\ei}{\end{itemize}}
    	\newcommand{\nn}{\nonumber}
            \newcommand{\r}[1]{(\ref{e:#1})}
            \newcommand{\edfl}[1]{\Label{#1}\end{df}}

\newcommand{\ve}{{\varepsilon}}

\newcommand{\dif}{{\partial}}
\newcommand{\half}{\frac{1}{2}}
	
	\newcommand{\ldif}{{\stackrel{\leftarrow}{\partial}}}
	\newcommand{\ldel}{{\stackrel{\leftarrow}{\delta}}}

\newcommand{\lG}{{\stackrel{\leftarrow}{G}}}

\newcommand{\lnabla}{{\stackrel{\leftarrow}{\nabla}}}

\begin{document}
\begin{titlepage}

\vspace*{5 mm}
\vspace*{20mm}
\begin{center}
{\LARGE\bf
Gauge theory}\end{center}\begin{center}{\LARGE\bf
of second class constraints}\end{center}
\begin{center}{\LARGE\bf without extra variables.}\end{center}
\vspace*{3 mm}
\begin{center}
\vspace*{3 mm}

\begin{center}Igor Batalin\footnote{On leave of absence from
P.N.Lebedev Physical Institute, 117924  Moscow, Russia\\E-mail:
batalin@td.lpi.ac.ru.} and Robert
Marnelius\footnote{E-mail: tferm@fy.chalmers.se}
 \\ \vspace*{7 mm} {\sl
Institute of Theoretical Physics\\ Chalmers University of Technology\\
G\"{o}teborg University\\
S-412 96  G\"{o}teborg, Sweden}\end{center}
\vspace*{25 mm}
\begin{abstract}
We show that any   theory with second class constraints may be cast into a gauge
theory if one makes use of  solutions of the constraints expressed in terms
of the
coordinates of the original phase space. We perform a Lagrangian path integral
quantization of the resulting gauge theory and show that the natural
measure follows
from a superfield formulation.
\end{abstract}\end{center}\end{titlepage}

\setcounter{page}{1}
\setcounter{equation}{0}
\section{Introduction.}
A covariant quantization of theories with second class constraints is in
general a
difficult task. A general method is the so called conversion method \cite{BF} in
which additional variables are added in such a way that the second
class constraints are converted into first class ones. This  allows then
for the use
of  conventional covariant quantization methods for general gauge theories.
In this
paper we show that another way to introduce additional variables, which
also enables
one to cast the original theory into a gauge theory, is to make use of
coordinates on
the constraint surface parametrized in terms of a redundant number of
variables. The
natural way to do this, which exactly yields the needed number of
variables, is to
choose the redundant variables to be   the coordinates of the enveloping,
original
phase space. No new additional variables are then added. This is the
procedure we
shall follow in this paper. If $x^i$ denotes the coordinates on the
original phase
space   and
$\theta^{\al}(x)=0$ the constraints, then we shall make use of functions
$\bx^i(x)$
satisfying the conditions
\be
&&\theta^{\al}(\bx^i(x))=0,
\e{1}
\be
&&\bx^i(\tx)=\tx^{i}
\e{2}
for whatever choice of solution $\tx^{i}$ of $\theta^{\al}(\tx)=0$. (The last
property  is a normalization of $\bx^i(x)$.) We shall show that $\bx^i(x)$
is gauge
invariant and that the original theory may be cast into a gauge theory
simply by replacing $x^i$ by
$\bx^i(x)$ in the (first order) action. Conventional gauge theoretical
quantization methods are then applicable.

The two conditions  above were previously considered in \cite{LM}. However,
there $\bx^i(x)$ were also required to satisfy
a closed Poisson algebra, which when combined with \r{1}, yields the condition
\be
&&\{\bx^i(x), \bx^j(x)\}=\{x^i,
x^j\}_D|_{x\ra\bx(x)},
\e{3}
where the bracket on the left-hand side is the original Poisson bracket
on the original phase space and where the bracket on the right-hand side is
the Dirac bracket. Solutions to these three conditions were then
investigated by means
of the general ansatz
\be
&&\bx^i(x)=x^i+\sum_{k=1}^\infty
X^i_{\al_1\cdots\al_k}(x)\theta^{\al_1}(x)\cdots\theta^{\al_k}(x).
\e{4}
This ansatz automatically satisfies \r{2} and it was shown that
 the coefficient functions
$X^i_{\al_1\cdots\al_k}(x)$  are possible to choose in many different ways to
solve the condition \r{1}. However, although the condition \r{3} is  more
restrictive a  general form for
the solutions of all three conditions were derived and a mapping procedure for
a covariant quantization by means of these solutions was proposed.

In this paper the philosophy is different. Here we  derive some general
properties
that only follow  from the existence of functions $\bx^i(x)$  satisfying
the first two
conditions,
\r{1} and
\r{2}. We  show that  the fact that  these solutions of the
constraints are expressed in terms of a redundant number of variables always
provide for the possibility of a gauge theoretical treatment of theories
with second
class constraints.

In section 2 we give the precise setting for our considerations.
In section 3 we present an auxiliary gauge theory and define some
properties needed
for our constructions. In section 4 we
demonstrate the general existence of $\bx^i(x)$ by means of an explicit integral
equation.  In section 5  we  prove that
$\bx^i(x)$ is gauge invariant  and in section 6 we show how the gauge
invariant action is constructed by means of $\bx^i(x)$ and  how it may be
quantized.
In section 7 we introduce a superfield formulation which is needed in the path
integral quantization to derive the appropriate measure  in a natural way.

\setcounter{equation}{0}
\section{The setting.}
Let $x^i$, $i=1,\ldots,2n$, be  coordinates on a symplectic supermanifold $\cM$,
$dim \cM=2n$, where $\ve_i\equiv\ve(x^i)$ are the Grassmann parities of
$x^i$. Let,
furthermore, there be a nondegenerate two-form
$\om$ on
$\cM$:
\be
&&\om=\half\om_{ij}(x)dx^j\wedge dx^i(-1)^{\ve_i},\quad {\rm sdet}
\,\om_{ij}\neq0,
\e{201}
which is required to be closed ($\dif_i=\dif/\dif x^i$):
\be
&&d\om=0\:\Leftrightarrow\:\dif_i\om_{jk}(x)(-1)^{(\ve_i+1)\ve_k}+cycle(i,j,
k)=0.
\e{202}
Since $\om$ is nondegenerate there exists an
inverse $\om^{ij}$ in terms of which the
super Poisson bracket is defined by
\be
&&\{A(x), B(x)\}=A(x)\ldif_i\om^{ij}(x)\stackrel{\ra}{\dif}_jB(x),
\nn\\&&\om^{ij}(x)\om_{jk}(x)=\om_{kj}(x)\om^{ji}(x)=\del^i_k.
\e{203}
$\om_{ij}(x)$ and $\om^{ij}(x)$ have the symmetry properties
($\ve(\om^{ij})=\ve(\om_{ij})=\ve_i+\ve_j$)
\be
&&\om_{ij}(x)=\om_{ji}(x)(-1)^{(\ve_i+1)(\ve_j+1)},\quad
\om^{ij}(x)=-\om^{ji}(x)(-1)^{\ve_i\ve_j}.
\e{2031}
The Poisson bracket \r{203} satisfies the Jacobi identities since \r{202}
implies
\be
&&\om^{ij}\ldif_{l}\om^{lk}(-1)^{\ve_i\ve_k}+cycle(ijk)=0.
\e{2032}

On $\cM$ we consider a Hamiltonian theory with the Hamiltonian $H(x)$.
Furthermore,
we let the theory be constrained by the conditions $\theta^{\al}(x)=0$,
 where  the Grassmann
parity of $\theta^{\al}$ is arbitrary and denoted by
$\ve_{\al}\equiv\ve(\theta^{\al})$.
$\theta^{\al}$ are linearly independent implying that the rank of
$\dif_i\theta^{\al}$ is equal to the number of constraints. (The rank
consists of two
blocks, one for the even part and one for the odd part. By the rank we mean
in the
following the sum of the two.) We are particularly interested in the case
when the
constraints are of second class. In this case we require the number of
constraints to
be $2m<2n$ and that
$\theta^{\al}$ satisfy
\be
&&\left.{\rm Rank}\;\{\theta^{\al}(x),
\theta^{\beta}(x)\}\right|_{\theta=0}=2m,
\e{205}
\be
&&\left.{\rm Rank}\;{\dif\theta^{\al}(x)\over\dif
x^i}\right|_{\theta=0}=2m.
\e{206}
 $\theta^{\al}(x)=0$
determines a constraint surface $\Gamma$ which in the case of second class
constraints
is  a symplectic supermanifold of dimension
$2(n-m)$.

\setcounter{equation}{0}
\section{An auxiliary gauge theory from general projection matrices}
Let us introduce the functions $Z^i_{\al}(x)$ satisfying the property
\be
&&\theta^{\beta}\ldif_iZ^i_{\al}=\del_{\al}^{\beta}, \quad
\ve(Z^i_{\al})=\ve_i+\ve_{\al}.
\e{600}
By means of $Z^i_{\al}$ we define the general projection matrices
$P_{\;j}^{i}$ by
\be
&&P_{\;j}^{i}(x)\equiv \del^i_j-Z^i_{\al}(x)\left(\theta^{\al}(x)\ldif_j\right).
\e{601}
$P_{\;j}^{i}$  satisfy then the following properties
\be
&&\theta^{\al}(x)\ldif_iP_{\;j}^{i}(x)=0,
\e{603}
\be
&&P_{\;j}^{i}(x)Z^j_{\al}(x)=0,
\e{604}
\be
&&P_{\;k}^{i}(x)P_{\;j}^{k}(x)=P_{\;j}^{i}(x).
\e{605}
From \r{600} it follows that the rank of $Z^i_{\al}$ is the same as the
rank of $\dif_i\theta^{\al}$. For
second class constraints we have therefore
\be
&& {\rm rank}\,Z^i_{\al}(x)=2m,\quad   {\rm rank}\,P^{i}_{\;j}(x)=2(n-m),
\e{606}
 from \r{206}.

By  means of $P^{i}_{\;j}(x)$ we may define the differential operator
\be
&&\lnabla_i\equiv \ldif_kP^{k}_{\;i}(x),
\e{607}
which due to \r{603} and \r{604} satisfies the properties
\be
&&\theta^{\al}(x)\lnabla_i=0,
\e{608}
\be
&&\lnabla_iZ^i_{\al}(x)=0.
\e{609}
As an additional condition on $Z^i_{\al}(x)$ we require that the differential
operators \r{607} satisfy the closed algebra
\be
&&[\lnabla_i, \lnabla_j]=\lnabla_kU_{ij}^k(x),
\e{610}
which partly is a consistency condition for \r{608}.   Notice that $U_{ij}^k(x)$
is not uniquely defined by this algebra
  since
\r{610} is unaffected by the replacement
$U_{ij}^k(x)\;\ra\;U_{ij}^k(x)+Z^k_{\al}(x)f_{ij}^{\al}(x)$ due to \r{609}.

The two conditions \r{600} and \r{610} on $Z^i_{\al}$ allow us to view the
differential operators
$\lnabla_i$  as gauge generators in a theory in which the constraint variables
$\theta^{\al}(x)$ are the physical gauge invariant variables. Any action
which only depends on
$\theta^{\al}(x)$ is then gauge invariant and the resulting gauge theory is
reducible due to \r{609}.
 From the prescription in \cite{BV} the  master action for this reducible gauge
theory is
 (using the short-handed DeWitt notation)
\be
&&S=\cS(\theta)+x_i^*P_k^{\;i}\cC^k+\half\cC_k^*U_{ij}^k\cC^j\cC^i(-1)^{\ve_i}+
\cC^*_iZ^i_{\al}\cC^{\al}_1-\cC^*_{1\al}U^{\al}_{\beta
i}\cC^i\cC^{\beta}_1(-1)^{\ve_{\beta}}+\nn\\&&+{1\over
6}\cC^*_{1\al}U^{\al}_{ijk}\cC^k\cC^j\cC^i(-1)^{\ve_i\ve_k+\ve_j},
\e{611}
where $\cS(\theta)$ is an arbitrary action only depending on
$\theta^{\al}(x)$, $\cC^i$
($\ve(\cC^i)=\ve_i+1$) are ghosts,
$\cC^{\al}_1$ ($\ve(\cC^{\al}_1)=\ve_{\al}$) ghosts for ghosts, and where
$x^*_i$, $\cC^*_i$, and
$\cC^*_{1\al}$ are antifields to $x^i$, $\cC^i$, and $\cC^{\al}_{1}$ with
opposite Grassmann parities to
the latter. The master equation $(S,S)=0$ contains all the previous
conditions. In fact, if we start from
the master action \r{611} and require $(S,S)=0$ for arbitrary actions
$\cS(\theta)$ all conditions are
generated.

\setcounter{equation}{0}
\section{An integral equation for the solution $\bx^i(x)$}
Let us introduce  the functions $\tx^i=\tx^i(\la,x)$, $i=1,\ldots,2n$,
where $\la$
is a bosonic parameter. Let furthermore this function satisfy the equation
\be
&&{d\tx^i\over d\la}=Z^i_{\al}(\tx)\theta^{\al}(\tx),\quad \tx^i(0,x)=x^i,
\e{101}
where
$Z^i_{\al}(x)$ is defined in \r{600}.
 This equation  implies
\be
&&{d\theta^{\al}(\tx)\over
d\la}=\theta^{\al}(\tx){\ldif\over\dif\tx^i}{d\tx^i\over
d\la}=\theta^{\al}(\tx)
\e{103}
due to \r{600}. Hence we have
\be
&&\theta^{\al}(\tx)=e^{\la}\theta^{\al}(x).
\e{104}
This implies in turn that
\be
&&\theta^{\al}(\bx)=0\quad{\rm
for}\quad\bx^i(x)\equiv\lim_{\la\ra-\infty}\tx^i(\la, x).
\e{105}
Since the equation \r{101} may always be solved, $\bx^i(x)$ always exists.
The equation \r{101} may be integrated to the following nonlinear Volterra
integral
equation
\be
&&\tx^i(\la, x)=x^i+\int_0^{\la} d\sigma
e^{\sigma}Z_{\al}^i(\tx(\sigma,x))\theta^{\al}(x),
\e{106}
where we have made use of \r{104}.
By means of iterations one may then obtain an expression of the form \r{4}
for $\bx^i(x)$ used in
\cite{LM} generalized to  coordinates with arbitrary Grassmann parities. To the
lowest orders in
$\theta^{\al}$ we get explicitly
\be
&&\bx^i(x)\equiv \lim_{\la\ra-\infty}\tx^i(\la,
x)=x^i-Z^i_{\al}(x)\theta^{\al}(x)+
\half Z^i_{\al}(x)\ldif_k Z^k_{\beta}(x)
\theta^{\beta}(x)\theta^{\al}(x)+\cdots\nn\\
\e{108}

The solution for $\bx^i(x)$ above imply that
\be
&&{\bx^i\ldif_k}=P^{i}_{\;m}(\bx)\sigma^{m}_{\;k}(x),
\e{302}
where
$\sigma^{m}_{\;k}(x)$ is an invertible
matrix function  normalized
 such that
$\left.\sigma^{m}_{\;k}(x)\right|_{\theta(x)=0}=\del^m_k$.
Notice, however,  that we may always
replace
$\sigma^{m}_{\;k}(x)$ by
\be
&&\sigma^{m}_{\;k}(x)\longrightarrow
\sigma^{m}_{\;k}(x)+Z_{\al}^m(\bx(x))M^{\al}_k(x)
\e{304}
without affecting  \r{302} due to \r{604}.
The expression \r{302}
satisfies the consistency condition
\be
&&0=\theta^{\al}(\bx)\ldif_k=\theta^{\al}(\bx){\ldif
\over\dif\bx^i}\left({\bx^i\ldif_k}\right)
\e{303}
due to \r{603}.
The integrability conditions ($\bx^i\ldif_{[k}\ldif_{l]}=0$) of
\r{302} may be written as
\be
&&P^i_{\;j}(\bx)\left(\sigma^{j}_{\;[k}(x)\ldif_{l]}-U_{nm}^j(\bx(x))
\sigma^m_{\;k}(x)\sigma^n_{\;l}(x)(-1)^{\ve_k\ve_n}\right)=0,
\e{305}
where we have required $Z^i_{\al}(x)$ also to satisfy \r{610}. Due to \r{604}
eq.\r{305} implies
\be
&&\sigma^{j}_{\;[k}(x)\ldif_{l]}=U_{nm}^j(\bx(x))
\sigma^m_{\;k}(x)\sigma^n_{\;l}(x)(-1)^{\ve_k\ve_n}+Z^j_{\al}\sigma^{\al}_{kl},
\e{306}
where $\sigma_{kl}^{\al}$ is antisymmetric in
the lower indices. Notice that antisymmetry here is meant in a super sense:
$[ij]=ij-ji(-1)^{\ve_i\ve_j}$. The relations \r{306} may equivalently be
written as
\be
&&(\sigma^{-1})^r_{\;[p}\ldif_k(\sigma^{-1})^k_{\;q]}=
(\sigma^{-1})^r_{\;j}U_{pq}^{j}(\bx)+
(\sigma^{-1})^r_{\;j}Z_{\al}^j(\bx)\sigma_{kl}^{\al}(\sigma^{-1})^l_{\;p}
(\sigma^{-1})^{k}_{\;q}(-1)^{\ve_p\ve_k}.\nn\\
\e{307}
Solutions of these conditions always exist since $\bx^i(x)$ exists.

\setcounter{equation}{0}
\section{Gauge invariance of $\bx^i(x)$}
The expression \r{302}  implies now
\be
&&\bx^i(x)\lG_{\al}(x)=0,\quad \ve(\lG_{\al})=\ve_{\al},
\e{401}
where
\be
&&\lG_{\al}(x)=\ldif_kG_{\al}^k(x),\quad
G_{\al}^k(x)\equiv (\sigma^{-1})^k_{\;m}(x)Z_{\al}^m(\bx(x)).
\e{402}
The integrability conditions \r{307} imply furthermore that the  algebra of
$G_{\al}$ is
closed:
\be
&&[\lG_{\al}(x), \lG_{\beta}(x)]=\lG_{\ga}(x)U_{\al\beta}^\ga(x),
\e{403}
where
\be
&&U_{\al\beta}^\ga(x)=V_{\al\beta}^{\ga}(\bx)-\sigma_{mn}^{\ga}(x)
(\sigma^{-1})_{\;l}^{n}(x)(\sigma^{-1})_{\;k}^{m}(x)
Z_{\al}^k(\bx)Z_{\beta}^l(\bx)(-1)^{\ve_l(\ve_m+\ve_{\al})},\nn\\
\e{404}
where in turn $V_{\al\beta}^{\ga}(\bx)$ is defined in \r{409} below and
 $\sigma_{ij}^{\ga}(x)$ in
\r{306}.\\
{\bf  Proof}: We have
\be
&&[\lG_{\al}(x), \lG_{\beta}(x)]=\ldif_iG^i_{[\al}\ldif_jG^j_{\beta]},\nn\\
&&G^i_{[\al}\ldif_jG^j_{\beta]}=(\sigma^{-1})^i_{\;m}(x)
(Z_{[\al}^m\lnabla_nZ_{\beta]}^n)(\bx)+\nn\\&&+
(\sigma^{-1})^{i}_{\;[m}(x)\ldif_j(\sigma^{-1})^{j}_{\;n]}(x)
Z_{\beta}^n(\bx)Z_{\al}^m(\bx)(-1)^{\ve_{\beta}(\ve_m+\ve_{\al})},
\e{405}
where we have made use of the expression \r{302} in the first term.
The first term is now zero due to \r{609}.
By means of the integrability conditions \r{307}
 we get then
\be
&&G^i_{[\al}\dif_jG^j_{\beta]}=-(\sigma^{-1})^{i}_{\;l}U^l_{jk}Z^k_{\al}
Z^j_{\beta}(-1)^{\ve_{\al}\ve_j}-\nn\\&&-(\sigma^{-1})^{i}_{\;j}Z^j_{\ga}(\bx)
\sigma_{mn}^{\ga}(x)
(\sigma^{-1})_{\;l}^{n}(x)(\sigma^{-1})_{\;k}^{m}(x)
Z_{\al}^k(\bx)Z_{\beta}^l(\bx)(-1)^{\ve_l(\ve_m+\ve_{\al})}.
\e{407}
Now
\be
&&P^i_{\;l}U_{jk}^lZ_{\al}^kZ_{\beta}^j(-1)^{\ve_{\al}\ve_j}
=P^i_{\;[j}\ldif_lP^l_{\;k]}Z_{\al}^k
Z_{\beta}^j(-1)^{\ve_{\al}\ve_j}=0
\e{408}
due to \r{604}. Hence, we must have
\be
&&U_{jk}^lZ_{\al}^kZ_{\beta}^j(-1)^{\ve_{\al}\ve_j}=-Z_{\ga}^l
V_{\al\beta}^{\ga}.
\e{409}
This inserted into \r{407} yields \r{403} with \r{404}.

\setcounter{equation}{0}
\section{Construction of gauge invariant actions and their quantization}
The action ($\la_{\al}$ are Lagrange multipliers)
\be
&&\cS[x]=\int dt \left(V_i(x)\dx^i-H(x)-\la_{\al}\theta^{\al}(x)\right)
\e{501}
describes  dynamics of the type we  consider although not in its most
general form
 since the equations of motion imply
\be
&&\om_{ij}=\dif_iV_j+\dif_jV_i(-1)^{(\ve_i+1)(\ve_j+1)},
\e{502}
which means that the
two-form
\r{201}  here is exact. (A general action may be written as in \cite{BF2}.)
The action \r{501} is in a first order form which according to a basic
theorem allows us to construct an
equivalent action by  replacing $x^i$ in \r{501} by the solutions
$\bx^i(x)$ of the constraints, \ie $\theta^{\al}(\bx)=0$. The equivalent
action is  then
\be
&&\cS[\bx]=\int dt \left(V_i(\bx)\dot{\bx}^i-H(\bx)\right).
\e{503}
This action  is now gauge invariant. We have
\be
&&\cS[\bx]{\ldel\over\del\bx^i(x)}G^i_{\al}(x)=0,
\e{512}
where $G^i_{\al}$ is given in \r{402}.
$\cS[\bx]$ may  be quantized by a Lagrangian path integral method (see
\cite{BV}). The
master action is
\be
&&S=\cS[\bx]+\int
dt\left(
x^*_iG^i_{\al}(x)\cC^{\al}+\half\cC^*_{\ga}U_{\al\beta}^{\ga}(x)
\cC^{\beta}\cC^{\al}(-1)^{\ve_\al}\right),
 \e{513}
where $\cC^{\al}$  are ghosts with Grassmann parity
$\ve(\cC^{\al})=\ve_{\al}+1$, and
$x^*_i$ and
$\cC^*_{\al}$ are antifields to
$x^i$ and
$\cC^{\al}$ with opposite Grassmann parities to the latter.
$U_{\al\beta}^{\ga}$ is
given by \r{404}. The master equation
$(S,S)=0$ is satisfied by the properties of $G^i_{\al}(x)$ in \r{402}.

\setcounter{equation}{0}
\section{A superspace formulation}
The path integral quantization of the gauge invariant action \r{503} using
the master action \r{513} does
not determine the natural measure. Here we demonstrate that this measure is
directly obtained if we make use  of a superfield formulation. We
follow then the particular formulation given in \cite{BBD,BBD2}. The
coordinates $x^i$
on the supersymplectic manifold are then turned into superfields according
to the
following rule:
\be
&&x^i\;\ra\; x^i(\tau)\equiv x^i_0+\tau x^i_1,\quad \ve(x^i(\tau))=\ve_i,
\e{701}
where $\tau$ is an odd Grassmann parameter. $x^i_0$ represents the original
coordinates $x^i$ with Grassmann parity $\ve_i$ while $x^i_1$ is the
superpartner to
$x^i_0$ with Grassmann parity $\ve(x^i_1)=\ve_i+1$. We may then define
superfunctions
$Z^i_{\al}(x(\tau))$ and $P^i_{\;j}(x(\tau))$ which are equal to the previously
considered functions with $x^i$ replaced by the superfields \r{701}. We may
therefore
also determine superfield solutions, $\bx^i(x(\tau))$, to the constraints
satisfying
\be
&&\theta^{\al}(\bx^i(x(\tau)))=0
\e{702}
along the lines of section 4. In fact, $\bx^i(x(\tau))$ are equal to the
solutions
\r{106},\r{108} with $x^i$ replaced by $x^i(\tau)$ which is easily seen by
expanding
\r{702} in $\tau$. These super solutions will then be gauge invariant in
the super
sense
\be
&&\bx^i(x(\tau))\lG_{\al}(\tau)=0,\quad
\lG_{\al}(\tau)\equiv\ldif_iG^i_{\al}(x(\tau)),
\e{703}
where $G^i_{\al}(x(\tau))$ is given by \r{402} with the replacement \r{701}.

Instead of the original action \r{501} we consider here the superfield action
\cite{BBD,BBD2}
\be
&&\cS'[x(\cdot)]\equiv \int dt d\tau
\left(V_i(x(\tau))Dx^i(\tau)(-1)^{\ve_i}-Q(x(\tau),
\tau)-\la_{\al}(\tau)\theta^{\al}(x(\tau))\right),\nn\\
\e{704}
where $Q$ is an odd function of the superfields \r{701} and $\tau$, and
where $V_i$
is the superpotential in \r{501} here expressed in terms of the superfield
\r{701}.
$\la_{\al}(\tau)$ is an independent superfield (Lagrange multiplier) with
Grassmann
parity $\ve(\la_{\al})=\ve_{\al}+1$.
$D$ is the odd differential operator
\be
&&D\equiv {d\over d\tau}+\tau {d\over dt},\;\;\Rightarrow \;\;D^2={d\over dt}.
\e{705}
A
variation of the action \r{704} yields the equations
\be
&&Dx^i(\tau)=-\{Q(\tau), x^i(\tau)\}-\la_{\al}(\tau)\{\theta^{\al}(x(\tau)),
x^i(\tau)\},\quad
\theta^{\al}(x(\tau))=0.
\e{706}
The consistency conditions
\be
&&0=D\theta^{\beta}(x(\tau))=-\{Q(\tau),
\theta^{\beta}(x(\tau))\}-\la_{\al}(\tau)\{\theta^{\al}(x(\tau)),
\theta^{\beta}(x(\tau))\}
\e{707}
determine $\la_{\al}$ in the case of second class constraints. The resulting
expression for
$\la_{\al}$  inserted back into
\r{706} yields then the equation
\be
&&Dx^i(\tau)=-\{Q(\tau),
x^i(\tau)\}_D,
\e{708}
where we make use of the Dirac bracket.
This  in turn implies by means of \r{705}
\be
&&\dx^i(\tau)=\{x^i(\tau),  H(x(\tau))\}_D, \quad
H(x(\tau))\equiv\dif_{\tau}Q(\tau)-\half\{Q(\tau),Q(\tau)\}_D.
\e{709}
This demonstrates how the equations from the superaction \r{704} reduces to
the equations from \r{501}
together with the equations for the superpartners.
If the theory allows for a
supersymmetric formulation, then $Q$ may be chosen to have no explicit
$\tau$-dependence.  In this case \r{704} is manifestly supersymmetric.

A gauge invariant superfield action is now obtained if one replaces
$x^i(\tau)$ in
\r{704} by the solutions of \r{702}, \ie $\bx^i(x(\tau))$. This action,
$\cS'[\bx]$,
may then be quantized using the superfield formulation in \cite{BBD,BBD2}.
Antibrackets and
$\Delta$-operators are defined by
\be
&&(F, G)\equiv F\int {\ldel\over\del\Phi^A(\tau, t)}d\tau
dt{\stackrel{\ra}{\del}\over\del\tphi_A(\tau, t)}G-(F\leftrightarrow
G)(-1)^{(\ve_F+1)(\ve_G+1)},\nn\\
&&\Delta\equiv -\int{\del\over\del\Phi^A(\tau, t)}d\tau
dt{{\del}\over\del\tphi_A(\tau, t)},
\e{710}
where $\tphi_A$ are super antifields with Grassmann parity
$\ve(\tphi_A)=\ve(\Phi^A)=\ve_A$. The functional derivatives satisfy
\be
&&{\del\over\del\Phi^A(\tau,
t)}\Phi^B(\tau',t')=\del_A^B\del(\tau-\tau')\del(t-t')=\Phi^B(\tau,t)
{\ldel\over\del\Phi^A(\tau',
t')},\nn\\
&&\ve\left({\del\over\del\Phi^A}\right)=
\ve\left({\ldel\over\del\Phi^A}\right)=\ve_A+1.
\e{711}
We use the conventions
\be
&&\int d\tau \tau=1,\quad \del(\tau-\tau')=\tau-\tau',\nn\\
&&\Rightarrow\quad \int f(\tau')\del(\tau-\tau')d\tau'=f(\tau)=\int
d\tau'\del(\tau'-\tau)f(\tau').
\e{712}
If the super field-antifields pairs, $\Phi^A$ and $\tphi_A$, in \r{710} are
defined as
follows
\be
&&\Phi^A(\tau)=\Phi_0^A+\tau\Phi_1^A,\nn\\
&&\tphi_A(\tau)\equiv \Phi_{1A}^*-\Phi_{0A}^*\tau,\quad
\ve(\Phi^A)=\ve(\Phi_0^A)\equiv\ve_A,
\e{713}
then the expressions \r{710} reduce to the conventional expressions in
terms of the
fields $\Phi^A_a$ and the corresponding antifields $\Phi^*_{aA}$,
$\ve(\Phi^*_{aA})=\ve_A+1$, ($a=0,1$). Notice that
\be
&&(\Phi^A(\tau,t), \tphi_B(\tau',t'))=-(\Phi_0^A(t),
\Phi_{0B}^*(t'))\tau'+\tau(\Phi_1^A(t),
\Phi_{1B}^*(t'))=\nn\\&&=\del^A_B\del(\tau-\tau')\del(t-t').
\e{7131}
The functional derivatives
satisfying
\r{711} are in terms of the component fields \r{713} given by
\be
&&{\del\over\del\Phi^A(\tau,t)}=\tau
{\del\over\del\Phi^A_0(t)}+(-1)^{\ve_A}{\del\over\del\Phi^A_1(t)},\quad
{\ldel\over\del\Phi^A(\tau,t)}=-
{\ldel\over\del\Phi^A_0(t)}\tau+{\ldel\over\del\Phi^A_1(t)},\nn\\
&&{\del\over\del\tphi_A(\tau,t)}=\tau
{\del\over\del\Phi^*_{1A}(t)}+{\del\over\del\Phi^*_{0A}(t)},\quad
{\ldel\over\del\tphi_A(\tau,t)}=-
{\ldel\over\del\Phi^*_{1A}(t)}\tau+(-1)^{\ve_A}{\ldel\over\del\Phi^*_{0A}(t)
}.\nn\\
\e{7132}

With these tools at hand we get the master action
\be
&&S=\cS'[\bx(\cdot)]+\int \tx_i(\tau, t) d\tau dt G^i_{\al}(x(\tau,
t))\cC^{\al}(\tau,
t)+\nn\\&&+\half \int \tca_{\ga}(\tau, t) d\tau dt
U_{\al\beta}^{\ga}(x(\tau,t))\cC^{\beta}(\tau, t)\cC^{\al}(\tau,
t)(-1)^{\ve_{\al}}-\nn\\&&-\int \tilde{\bar{\cC^{\al}}}(\tau, t) d\tau dt
\la_{\al}(\tau, t)(-1)^{\ve_{\al}},\nn\\
\e{714}
where $\cS'[\bx(\cdot)]$ is the action \r{704} with $x^i(\tau)$ replaced by
$\bx^i(x(\tau))$ in
\r{702} and where the last term is a standard nonminimal term to allow for
a gauge
fixing delta function. Notice that $\cS'[\bx(\cdot)]$ does not contain
$\la_{\al}$ in
\r{704} and that $\la_{\al}$ in the last term is  a new variable. The master
action
\r{714} satisfies
\be
&&(S,S)=0,\quad \Delta S=0,
\e{715}
where the antibracket and the $\Delta$-operator are given by \r{710} for
the set of
superfields
$\Phi^A=\{x^i, \cC^{\al}, \bar{\cC}^{\al}, \la_{\al}\}$ and their super
antifields. Note that $\ve(\cC^{\al})=\ve_{\al}+1$ and
$\ve(\bar{\cC}^{\al})=\ve_{\al}$. The second equality in
\r{715} follows from the locality in
$\tau$ of
$S$ which yields a factor zero and implies that $S$ also satisfies the
quantum master
equation which in turn implies that no quantum corrections of the natural
measure in
the path integral is required. In order to gauge fix the master action
\r{714} we need a gauge fixing fermion, $\Psi$, expressed in terms of the
superfields
$\Phi^A$ such that the super antifields are determined through the equations
\be
&&\tphi_A(\tau, t)=\Psi{\ldel\over\del\Phi^A(\tau,
t)}=(-1)^{\ve_A}{\del\over\del\Phi^A(\tau, t)}\Psi.
\e{716}
A possible choice is
\be
&&\Psi=\int \theta^{\al}(x(\tau, t)) d\tau dt \bar{\cC}_{\al}(\tau, t)=\int
d\tau dt
\bar{\cC}_{\al}(\tau, t)\theta^{\al}(x(\tau, t)).
\e{717}
Eq.\r{716} yields then
\be
&&\tx_i(\tau, t)=\bar{\cC}_{\al}(\tau, t)\theta^{\al}(x(\tau,
t))\ldif_i(-1)^{\ve_i},\quad
\tilde{\cC}_{\al}(\tau, t)=0,\nn\\&& \tilde{\la}^{\al}(\tau, t)=0,\quad
\tilde{\bar{\cC^{\al}}}(\tau, t)=\theta^{\al}(x(\tau, t)).\nn\\
\e{718}
With these expressions inserted into \r{714} we obtain the
 gauged fixed action
\be
&&S_{\Psi}=\cS'[\bx(\cdot)]+\int d\tau dt
\bar{\cC}_{\al}(\tau, t)\theta^{\al}(x(\tau, t))\ldif_iG^i_{\beta}(x(\tau,
t))\cC^{\beta}(\tau, t)-\nn\\&&-\int d\tau dt\la_{\al}(\tau,
t)\theta^{\al}(x(\tau,
t)).
\e{719}
In the path integral the last term yields the delta-function
$\del(\theta^{\al})$ after integration over $\la_{\al}$. In the presence of
this delta-function $\bx^i(x)=x^i$, and the
middle term yields unity after integration over $\cC^{\al}$ and
$\bar{\cC}_{\al}$
since
$\left.G^i_{\beta}\right|_{\theta=0}=Z^i_{\beta}$ implies
$\left.\theta^{\al}\ldif_iG^i_{\beta}\right|_{\theta=0}=\del^{\al}_{\beta}$
due to
\r{600}. Thus, the path integral over $S_{\Psi}$ reduces to the path
integral over
the first and last term in \r{719} which is equivalent to a path integral
over the
original action
\r{704}. Integration over the superpartner
$x_1^i$ in \r{701} yields then the expected measure as was shown in
\cite{BBD2}.\\

\noindent
{\bf Acknowledgements}:

I.A.B. would like to thank Lars Brink for
his very warm hospitality at the
Department of Theoretical Physics, Chalmers
and G\"oteborg University.
 The work of I.A.B. is supported by the grants 99-01-00980 and
99-02-17916 from Russian foundation for basic researches and by the
President grant
00-15-96566 for supporting leading scientific schools. The work is
partially supported
by INTAS grant 00-00262.

\end{document}